\documentstyle[preprint,aps,epsfig,floats,amssymb]{revtex}
\begin{document}

\preprint{
\vbox{
\hbox{January 2001}
\hbox{ADP-01-02/T437}
}}

\title{Chiral Symmetry and the Intrinsic Structure of the Nucleon}

\author{D.B.~Leinweber, A.W.~Thomas and R.D.~Young}
\address{	Special Research Centre for the
		Subatomic Structure of Matter,
		and Department of Physics and Mathematical Physics,
		Adelaide University, Adelaide SA 5005,
		Australia}

\maketitle

\begin{abstract}
Understanding hadron structure within the framework of QCD is an
extremely challenging problem.  In order to solve it, it is vital that
our thinking should be guided by the best available insight.  Our
purpose here is to explain the model independent consequences of the
approximate chiral symmetry of QCD for two famous results concerning
the structure of the nucleon.  We show that both the apparent success
of the constituent quark model in reproducing the ratio of the proton
to neutron magnetic moments and the apparent success of the Foldy term
in reproducing the observed charge radius of the neutron are
coincidental.  That is, a relatively small change of the current quark
mass would spoil both results.
\end{abstract}

\newpage

The chiral properties of QCD have been the subject of considerable
attention, from chiral quark models\cite{Manohar:1984md,Brown:1979ui,CBM}
to the less ambitious, but more
systematic, approach of chiral perturbation theory\cite{Chpth}.
Most recently one has begun to
realize the importance of chiral symmetry in describing the dependence of
hadron properties like masses\cite{MASSES}
and magnetic moments\cite{MAGMOM} on quark mass.  This
is vital if one is to compare lattice QCD calculations, which are
presently confined to current quark masses, $\bar{m}$, of 
order 40-80 MeV or higher, with experimental data.

For our purposes the essential point is that chiral symmetry is
dynamically broken. The resulting Goldstone bosons
enter the calculation of hadron properties through loops which lead
to a characteristic dependence on $\bar{m}$ which is not analytic.  Indeed for
the magnetic moment of the nucleons one finds a leading non-analytic 
behavior proportional to $m_q^{1/2}$. In the chiral limit 
$m_\pi^2\propto m_q$ and
\begin{eqnarray}
\mu^p &=& \mu^p_0 - \alpha m_\pi + {\cal O}(m_\pi^2), 
\nonumber \\
\mu^n &=& \mu^n_0 + \alpha m_\pi + {\cal O}(m_\pi^2).
\label{eq:muchiral}
\end{eqnarray}
It is a crucial property of the leading non analytic (LNA) coefficient,
$\alpha$, 
that it is entirely determined by the axial charge of the nucleon and
the pion decay constant (both in the chiral limit):
\begin{equation}
\alpha = \frac{g_A^2 M_N}{8 \pi f_\pi^2}.
\label{eq:alpha}
\end{equation}
Taking the one-loop value of $g_A (= F_1 + D_1 = 0.40 + 0.61)$ from chiral
perturbation theory\cite{jenkins} we find $\alpha =4.41$.
(Note that all magnetic moments
will be in nuclear magnetons ($\mu_N$) and all masses in GeV).

Clearly the LNA term is large, of order 0.6 $\mu_N$, 
at the physical pion mass.  This is
one third of the magnetic moment of the neutron. 
Provided the ${\cal O} (m_\pi^2)$ terms are small at the
physical pion mass we can use Eq.(\ref{eq:muchiral}) 
to extract the proton and neutron
magnetic moments in the chiral limit:
\begin{eqnarray}
\mu^p_0 &\cong& \mu^p + \alpha m_\pi^{\rm phys}, 
\nonumber \\
\mu^n_0 &\cong& \mu^n - \alpha m_\pi^{\rm phys}.
\label{eq:mu0}
\end{eqnarray}
One then finds a model independent expression for the dependence of the proton
to neutron magnetic moment ratio on the pion mass: 
\begin{equation}
\frac{\mu^p}{|\mu^n|}  =  \frac{\mu^p_0}{|\mu^n_0|} 
\left( 1 + \left[ \frac{1}{|\mu^n_0|} - \frac{1}{|\mu^p_0|} \right]
\alpha m_\pi \right) + {\cal O} (m_\pi^2).
\label{eq:ratio}
\end{equation}
Constraining the chiral expansions to reproduce the experimental 
proton moment $\mu^p$ and the experimental ratio $\mu^p/|\mu^n|$ provides
\begin{equation}
\mu^p_0 = 3.41\ \mu_N, \hspace{10mm} \frac{\mu^p_0}{|\mu^n_0|} = 1.37,
\label{eq:mu0p}
\end{equation}
and
\begin{equation}
\frac{\mu^p}{|\mu^n|} = 1.37 + 0.09 \frac{m_\pi}{m_\pi^{\rm phys}} + 
{\cal O} (m_\pi^2).
\label{eq:numratio}
\end{equation}

As a consequence of Eq.(\ref{eq:numratio}),
we see that the ratio of the $p$ to $n$ magnetic moments varies
from 1.37 to 1.55 (a variation of order 13\%) as $m_\pi$ varies from 
0 to $2m_\pi^{\rm phys}$.
In terms of the underlying quark mass, such a variation corresponds to a
current quark mass variation from 0 to just 20 MeV.  Within the 
constituent quark model 
this ratio would remain constant at 3/2, independent of the change of
quark mass.

A study by Leinweber et al. \cite{MAGMOM} suggests a model independent 
method for describing the mass dependence of baryon magnetic moments 
which satisfies the chiral
constraints imposed by QCD.  We briefly summarize the main results of
that analysis. A series expansion of $\mu _{p(n)}$ in powers of $m_{\pi }$ is
not a valid approximation for $m_{\pi }$ larger than the physical
mass. On the other hand, the simple Pad\'e approximant
\begin{equation}
\mu^{p(n)} = \frac{\mu^{p(n)}_0}{1 \pm \frac{\alpha}{\mu^{p(n)}_0} m_\pi +
\beta^{p(n) }\, m_\pi^2},
\label{eq:pade}
\end{equation}
has the correct leading non-analytic
(LNA) behavior of chiral perturbation theory
\[
\mu^{p(n)} =\mu_{0}^{p(n)} \mp \alpha\, m_{\pi} ,
\]
and also builds in the 
expected behavior at large $m_{\pi }$. At heavy quark masses we
expect that the magnetic moment should fall off as the Dirac moment
\[
\mu = \frac{e_q}{2m_q}\propto\frac{1}{m^2_{\pi }}
\]
as $m_{\pi }$ becomes moderately large.\footnote{Note that this form 
is valid provided $m_\pi^2 \propto \bar{m}$, which seems to be 
true for $m_\pi$ up to at least 1 GeV within lattice simulations.}
A fit of the Pad\'e approximant to lattice QCD data \cite{dblOctet,wilcox92} 
leads to predictions of the magnetic moments of 2.90(20) 
and $-1.79(21)\ \mu_N$ to be compared with 2.793 and $-1.193\ \mu_N$ for 
$p$ and $n$ respectively.

\begin{figure}[t]
\begin{center}
\epsfig{file=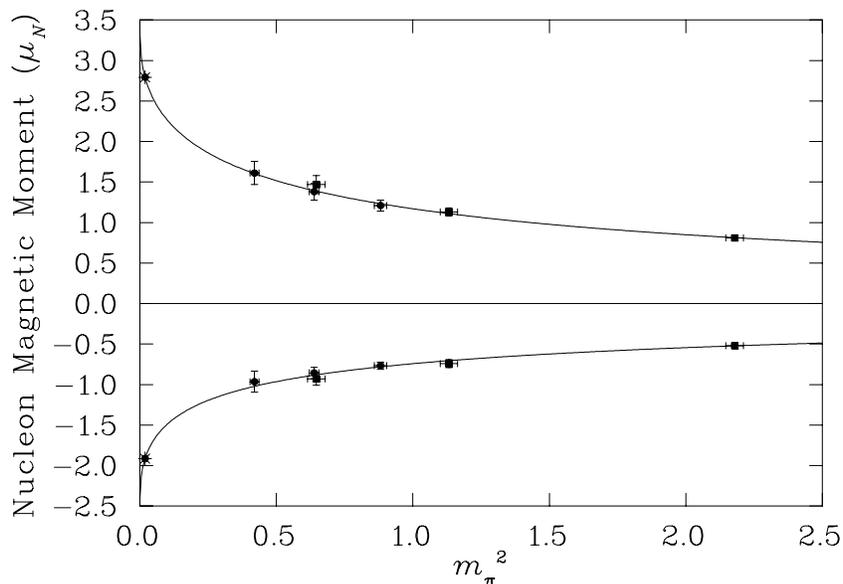, angle=90, width=11cm}
\caption{Extrapolation of lattice QCD magnetic moments 
($\bullet$ LDW Ref.\ \protect\cite{dblOctet},
$\scriptstyle\blacksquare$ WDL Ref.\ \protect\cite{wilcox92}) 
for the proton (upper) and neutron (lower) 
to the chiral limit. The curves are constrained to pass through the 
experimentally measured moments which are indicated by asterisks.
\label{fig:1}}
\end{center}
\end{figure}

\begin{figure}[t]
\begin{center}
\epsfig{file=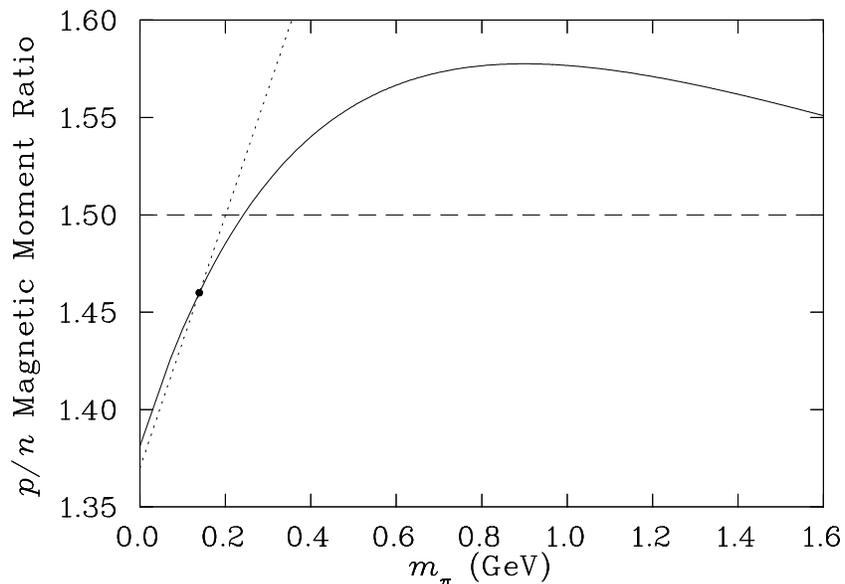, angle=90, width=11cm}
\caption{Ratio of the magnitudes of the proton to neutron magnetic moments. 
The solid curve describes the predictions of the Pad\'e approximant while 
the dashed line denotes the constituent quark model prediction of 3/2. 
The dotted line is the leading non-analytic behavior of chiral perturbation 
theory. The experimental measurement is indicated by the solid point.
\label{fig:2}}
\end{center}
\end{figure}

Figure \ref{fig:1} shows a similar fit to the lattice data, this time 
constrained to pass through the experimental moments, and providing 
the solid curve in Fig.~\ref{fig:2} for the $p/n$ ratio of magnetic 
moments.
The Pad\'e approximate fit parameters are 
$(\mu_0,\beta) = (3.33,0.527)$ and $(-2.41,0.427)$, for 
$p$ and $n$ respectively.
{}Figure \ref{fig:2} also shows the the result of the constituent
quark model (dashed line) and the variation of the ratio predicted by
the leading non-analytic behavior of chiral perturbation theory in
Eq.~(\ref{eq:numratio}) (dotted line).  The importance of the terms of
order $m_\pi^2$ and higher are revealed by the ratio calculated using
the Pad\'e approximant of Eq.(\ref{eq:pade}) (solid curve).  The
values of $\mu_0^{p(n)}$ vary slightly in the chiral expansion and the
Pad\'e due to these small higher order corrections at the physical
pion mass. However, it is important to note that the slopes of the
curves agree exactly in the chiral limit, as demanded by chiral
perturbation theory.

The key point is that the ratio displays a significant quark mass 
dependence. It is roughly linear in $m_\pi$ until $m_\pi$ is of order 
$2 m_\pi^{\rm phys}$. It is amusing to imagine the excitement had 
the pion mass been 100 MeV heavier at 240 MeV where the Pad\'e 
crosses the constituent quark model prediction of 3/2. However the 
constituent quark model prediction really 
corresponds to the $m_\pi\rightarrow\infty$ limit, and Fig.~\ref{fig:2} 
suggests this limit is approached rather slowly.

The surprising consequences of chiral symmetry for this famous ratio
naturally leads us to reconsider the neutron charge radius. The squared 
charge radius of the neutron ($<r^2>^n_{\rm ch}$) 
is obtained from the slope of the
neutron electric form factor, $G_{En}(Q^2)$ as $Q^2 \rightarrow 0$:
\begin{equation}
<r^2>^n_{\rm ch} = -6 \frac{d}{d Q^2} G_{E}(Q^2)|_{Q^2=0}
\label{eq:chradiusA}
\end{equation}
The Sachs electric and magnetic form factors can be written in terms
of the covariant vertex functions $F_1$ and $F_2$ as
\begin{eqnarray}
G_E(Q^2) & = & F_1(Q^2) - \frac{Q^2}{4 M_N^2} F_2(Q^2) \nonumber \\
G_M(Q^2) & = & F_1(Q^2) + F_2(Q^2).
\label{sachs}
\end{eqnarray}
Note that for a neutral charge particle $F_1(Q^2=0)$ vanishes and hence
$F_2(Q^2=0)$ is simply the magnetic moment of the particle. Now the 
charge radius squared of the neutron can be written as
\begin{equation}
<r^2>^n_{\rm ch} = -6 \frac{d}{d Q^2} F_{1n}(Q^2)|_{Q^2=0} 
+ \frac{3}{2}\frac{\mu^n}{M_N^2}.
\label{eq:chradius}
\end{equation}
Experimentally $<r^2>^n_{\rm ch} = -0.113 \pm 0.003 \pm 
0.004$ fm$^2$ \cite{expt},
while the last term in Eq.(\ref{eq:chradius}), the Foldy 
term\cite{Foldy}, is numerically 
$-$0.126 fm$^2$.

The close agreement between the Foldy term and the observed
mean square charge radius of the neutron has led to considerable
controversy.  It has been argued that the difference, namely the term
involving the Dirac form factor ($F_{1n}$), should be interpreted as 
the true indication of the intrinsic charge distribution 
of the neutron.  Clearly
this would be quite insignificant.  On the other hand, decades of
modeling the structure of the nucleon have suggested that the neutron
must have a non-trivial intrinsic charge distribution.  Pre-QCD it was
clear that the long-range tail must be negative, corresponding to the
emission of a negative pion ($n \rightarrow p \pi^-$), 
but old-fashioned meson theory was
incapable of describing the interior of the neutron.  Post-QCD this was
resolved in the cloudy bag model\cite{CBM,CBMn}, 
where the convergence of an expansion
in numbers of pions was assured --- provided the quark confinement region
was fairly large and the decuplet states (in this case the $\Delta(1232)$) 
was included on the same footing as the nucleon\cite{Dodd}. 
The neutron charge distribution
then originated mainly from the Fock component of its wave function
consisting of a $\pi^-$ cloud and a positive core of confined quarks.
Alternatively, within the constituent quark model, 
it was proposed that the repulsive gluon
exchange interaction between the two d-quarks would tend to force them
to the exterior of the neutron --- again yielding a positive core and a
negative tail\cite{IK}.

In view of these expectations of an internal charge distribution, the
interpretation of $<r^2>^n_{\rm ch}$ 
in terms of the Foldy term has
been controversial.  Isgur has recently shown that a careful treatment
of relativistic corrections for the calculation of $<r^2>^n_{\rm ch}$, in a
quark-di-quark model, leads to a recoil contribution that cancels
the Foldy term exactly, hence restoring the interpretation in terms
of an intrinsic charge distribution --- see also~\cite{coon}.  We now show 
that the study of the
chiral behavior of $<r^2>^n_{\rm ch}$ and $\mu^n$ supports this idea, 
establishing in a model
independent way that the observed similarity between the experimental
value and the Foldy term is purely accidental.

It is a little appreciated consequence of the approximate chiral symmetry of
QCD that the mean square charge radius of the nucleon has a leading
non-analytic term proportional to $\ln m_\pi$\cite{RADII}:
\begin{equation}
<r^2>^{p(n)}_{\rm ch}|_{\rm LNA} = \mp 
\frac{1 + 5 g_A^2}{(4 \pi f_\pi)^2} \ln \left( \frac{m_\pi}{\Lambda} \right),
\label{eq:LNA}
\end{equation}
where the upper and lower sign correspond to $p$ and $n$ respectively.
As a result, the charge radii of both $p$ and $n$ diverge
logarithmically as the quark mass tends to zero.  Physically this is
easy to understand; as $m_\pi \rightarrow 0$ the Heisenberg
Uncertainty Principle allows the pion cloud, and therefore the charge
density, to extend to infinite distance.  For the magnetic moment, on
the other hand, there is no divergence --- indeed the neutron magnetic
moment increases in magnitude by about 30\% as the pion mass moves
from its physical value to zero. (Loosely speaking, even though the
pion may be at a large distance it moves slowly; its angular momentum
is constrained to one by angular momentum conservation.)

To summarize, whereas a change of order 5 MeV in the light quark mass
leads to a 30\% change in the Foldy term, the neutron charge radius 
$<r^2>^n_{\rm ch}$ becomes infinite! Hence, the similarity of 
$<r^2>^n_{\rm ch}$ and the Foldy term is purely an accident. A small 
change in the quark mass leads to completely different values.
This physics is not captured in the constituent quark model where a 5 MeV 
change in the light quark mass corresponds to a change in the 
constituent quark mass from roughly 340 to 335 MeV. In this case the 
neutron charge radius originates in the one-gluon-exchange interaction 
which is proportional to the inverse square of the constituent quark 
mass and therefore $<r^2>^n_{\rm ch}$ would change by only 3\%.

In summary, chiral perturbation theory provides model independent 
constraints on the 
quark mass dependence of nucleon magnetic moments and charge radii 
which compel one to conclude that the apparent success of the 
constituent quark model 
to predict the $p/n$ magnetic moment is accidental. Had the pion mass 
been lighter than the observed value, the $p/n$ ratio would drop further 
from the constituent quark model prediction of 3/2; the latter 
corresponding to the 
$m_\pi\rightarrow\infty$ limit. The coincidence of the Foldy term 
and the observed neutron charge radius is also accidental. Here 
a small change in the quark mass to the chiral limit increases 
the neutron moment by about 30\% while the charge radius becomes infinite. 
These results, which are a rigorous consequence of the chiral symmetry 
of QCD, cannot be simulated in conventional constituent quark models.

\acknowledgements

This work was supported by the Australian Research Council.


\end{document}